# Adaptive Proton Therapy Using CBCT-Guided Digital Twins

Chih-Wei Chang[1*], Zhen Tian[2], Richard L.J. Qiu[1], H. Scott McGinnis[1], Duncan Bohannon[1], Pretesh Patel, Yinan Wang[1], David S. Yu[1], Sagar A. Patel[1], Jun Zhou[1*] and Xiaofeng Yang[1,*]

[1]Department of Radiation Oncology and Winship Cancer Institute, Emory University, Atlanta, GA 30308

[2]Department of Radiation and Cellular Oncology, University of Chicago, Chicago, IL, 60637

*Co-corresponding authors: chih-wei.chang@emory.edu (CC), jun.zhou@emory.edu (JZ), xiaofeng.yang@emory.edu (XY)

**Abstract**

**Objective:** This study aims to develop a digital twin (DT) framework to enhance adaptive proton stereotactic body radiation therapy (SBRT) for prostate cancer. Prostate SBRT has emerged as a leading option for external beam radiotherapy due to its effectiveness and reduced treatment duration. However, interfractional anatomy variations can impact treatment outcomes. This study seeks to address these uncertainties using DT concept, with the goal of improving treatment quality, potentially revolutionizing prostate radiotherapy to offer personalized treatment solutions.

**Approach:** A retrospective study on two-fraction prostate proton SBRT was conducted, involving a cohort of 10 randomly patient cases from an institutional database (n=43). DT-based treatment plans were developed using patient-specific clinical target volume (CTV) setup uncertainty, determined through machine learning predictions. Plans were optimized using pre-treatment CT and corrected cone-beam CT (cCBCT). Plan evaluation was performed using cCBCT to account for actual patient anatomy. The ProKnow scoring system was adapted to determine the optimal treatment delivery plans.

**Main Results:** The average CTV D98 values for original clinical and DT-based plans across 10 patients were 99.0% and 98.8%, with hot spots measuring 106.0% and 105.1%. Regarding bladder metrics, the clinical plans yielded average bladder neck V100 values of 29.6% and bladder V20.8Gy values of 12.0cc, whereas the DT-based plans show better sparing of bladder neck with the values of 14.0% and 9.5cc, respectively. Compared to clinical plans, the proposed DT-based plans enhance dosimetry quality, improving plan scores ranging from 2.0 to 15.5.

**Significance:** Our study presented a pioneering approach that leverages DT technology to enhance adaptive proton SBRT. The framework improves treatment plans by utilizing patient-specific CTV setup uncertainty, which is usually smaller than conventional clinical setups. This research contributes to the ongoing efforts to enhance the efficiency and efficacy of prostate radiotherapy, with ultimate goals of improving patient outcomes and life quality.

# 1 Introduction

Ultra-hypofractionated radiotherapy, involving the delivery of radiation doses greater than 5 Gy per fraction, has emerged as the standard of care for prostate cancer external beam radiation therapy (RT) with low rates of side-effects (Tree *et al.*, 2022; Ma *et al.*, 2024). This approach offers significant advantages over conventional RT regimens which typically involve smaller daily fractions administered over a more extended treatment course. Ultra-hypofractionation reduces the overall treatment time, providing greater convenience and cost-effectiveness for patients, and exploits the radiobiological principles of hypofractionation, potentially enhancing the therapeutic ratio (Hannan *et al.*, 2016; Kishan *et al.*, 2019). The advent of proton pencil beam stereotactic body RT (SBRT) has further revolutionized the delivery of ultra-hypofractionated RT for prostate cancer (Bryant *et al.*, 2016). This cutting-edge technique combines robust treatment planning (Liu *et al.*, 2012) with advanced image-guided systems (Zwart *et al.*, 2022), enabling the precise administration of ultra-high radiation doses within five fractions. By leveraging the unique physical properties of protons, proton SBRT can potentially offer the unparalleled capability for sparing healthy tissues compared to conventional photon-based modalities.

As emerging clinical data continue to validate the effectiveness and safety of prostate SBRT management, efforts are underway to further optimize treatment quality and outcomes. Two critical factors that can significantly enhance the therapeutic potential of prostate SBRT (Mancosu *et al.*, 2016) are (1) mitigating geometrical uncertainties and (2) maximizing dose conformity to targets while sparing the surrounding organs at risk (OARs). Geometrical uncertainties, arising from variations in anatomy between treatment fractions and positional inaccuracies, can compromise target coverage and increase the risk of toxicity. Adaptive radiotherapy strategies (Bobić *et al.*, 2021; Paganetti *et al.*, 2021), which involve modifying the treatment plan based on daily imaging, offer a promising solution for these uncertainties. By continuously adapting the plan to the patient's anatomy, target coverage can be maintained while minimizing unnecessary doses for healthy tissues. However, the current treatment planning techniques are limited by computational constraints, making real-time adaptive proton therapy challenging (Chang *et al.*, 2024). This work aims to investigate an image-guided digital twin (DT) framework to integrate patient-specific clinical target volume (CTV) positional setups with uncertainty and proton robust plan optimization. The goal is to create multiple digital replicants of treatment plans, enabling online decision-making for selecting the optimal treatment plan on the actual treatment day.

NASA (Glaessgen and Stargel, 2012) defined the concept of DT as integrating multiphysics, multiscale, probabilistic simulations to support decision-making during aeronautical space missions and manufacturing processes. Over time, the scope of DT has since been extended beyond replicating physical objects to encompass diagnosis, prognosis of system hazards, and optimization of operations (Rosen *et al.*, 2015). A digital twin concept framework (Kapteyn *et al.*, 2021) leverages multiphysics models to forecast event progression and employs data assimilation techniques to accurately evaluate system states, thereby delivering recommendations for appropriate control actions. This framework has been explored as an *in-silico* patient surrogate for personalized medicine (Björnsson *et al.*, 2019; Hormuth *et al.*, 2021), computationally treating and systematically evaluating all relevant practical regimens to identify the optimal treatment solution for an individual patient. Nonetheless, diagnostic and therapeutic decisions necessitate consideration of multiple factors (Wu *et al.*, 2022), such as symptoms, pathogenesis, and biopharmaceuticals, which may preclude the existence of a proper mechanism-based mathematical model to support the development of a digital twin framework in medicine. In this study, we narrowed the concept of digital twins to radiotherapy and formulated a case study to demonstrate its application in radiation treatment planning and delivery.

The proposed DT framework seeks to address the limitations of current adaptive proton therapy techniques by leveraging advanced computational methods and imaging data. By quantifying the uncertainties associated with the patient-specific CTV setup uncertainty and generating robust proton treatment plans tailored to potential anatomical variations, this approach aims to facilitate plan adaptation based on the patient's daily imaging data. Through the creation of multiple digital replicants of treatment plans, the framework empowers clinicians to make informed decisions regarding the most appropriate plan for the patient's anatomy on the day of treatment. This approach can potentially enhance target coverage, minimize dose to OARs, and streamline the adaptive radiotherapy workflow, potentially reducing treatment times and improving overall efficiency. According to the authors' best knowledge, this research pioneers the application of DT in external beam radiotherapy. The primary contributions of this study can be categorized into two distinct dimensions, both of which are geared toward assessing the feasibility of clinical implementation:

- The proposed framework demonstrates how to leverage the concept of digital twins to create multiple patient-specific prostate SBRT plans, enabling patients to benefit from proton adaptive therapy over the conventional RT, allowing for better dose conformity to the target while sparing healthy tissues.
- The proposed framework demonstrates how to assimilate computed tomography (CT) and corrected cone-beam CT (CBCT) data from prior treatments to predict the most likely patient-specific CTV positional setups with uncertainty, thereby reducing the robust optimization margins from institutional clinical guidelines.

## 2 Materials and methods

### 2.1 Patient data

We utilized institutional CBCT image data to investigate treatment planning and delivery using the proposed DT framework for adaptive proton therapy. The CBCT images served three purposes: predicting patient-specific setup uncertainty for CTV, robust proton treatment planning, and treatment evaluation. The CBCT image sets were retrieved from 43 prostate cancer patients who received five-fraction SBRT, with six of those patients receiving a simultaneously integrated boost (SIB) for a dominant intraprostatic lesion (DIL). Each of the 43 patients had five CBCT scans acquired on the Varian ProBeam® on-board imaging system, resulting in a total of 215 CBCT image sets. We randomly selected 10 patients, including 2 patients with DIL SIB, from the institutional prostate SBRT database to demonstrate the proposed DT framework. The remaining patients were employed to train a machine learning model for predicting patient-specific CTV setup uncertainty. Additionally, we included CBCT images from another group of 49 prostate cancer patients who underwent conventional 28-fraction proton therapy to expand the image database and provide prior knowledge for inferring the underlying correlation between CBCT images and the CTV. Each of these 49 patients had daily CBCT imaging during treatment, contributing 1,372 CBCT image sets. All patients had pre-treatment CT simulation images obtained from a Siemens SOMATOM Definition Edge scanner for initial treatment planning.

### 2.2 Treatment planning and two-fraction prostate SBRT

RayStation 2023B (RaySearch Lab., Stockholm, Sweden), was used to provide fast robust proton treatment planning and plan evaluation based on the same-day CBCT to enable real-time decision-making for delivering the optimal treatment plan. The treatment planning system (TPS) is deployed on a clinical GPU server with a single NVIDIA Quadro RTX 8000 and dual Intel® Xeon® Gold 6136 CPU. RayStation 2023B supports GPU-based deformable image registration and proton Monte Carlo (MC) dose calculation that can achieve online CBCT-based treatment evaluation in 2 minutes (Chang *et al.*, 2023b). At our institution, prostate SBRT plans were optimized using four proton beams, including bi-lateral, left anterior oblique (LAO), and right anterior oblique (RAO). The use of anterior-oblique beams can potentially limit the rectum doses (Moteabbed *et al.*, 2017). The beam model embedded a constant relative biological effectiveness (RBE) of 1.1 (IAEA/ICRU, 2008). Robust optimization was used for all clinical plans with 5 mm positional uncertainty (except 3 mm for posterior) and ±3.5% range uncertainty, resulting in 21 scenarios in each plan optimization.

This work utilizes data from the two-fraction prostate SBRT clinical trials, 2STAR (NCT02031328) (Alayed *et al.*, 2019) and 2SMART (NCT03588819) (Ong *et al.*, 2023a), to explore the feasibility of applying the proposed DT framework for adaptive proton therapy planning. For the 2STATR trial, the prescribed dose was 26 Gy delivered in two fractions to the CTV, with a one-week break between fractions. The 2SMART trial included an additional dose-escalated boost up to 32 Gy to the gross tumor volume (GTV). Table 1 shows the dose constraints for the organs at risk (OARs), including the bladder and rectum. Slightly higher dose limits were allowed for the bladder and rectum in the 2SMART trial to accommodate the GTV boost (Ong *et al.*, 2023b). Additionally, we contoured the bladder neck in this study, as it has been identified as strongly correlated with bladder toxicity (Hathout *et al.*, 2014). While there are no established dose constraints for the bladder neck, we utilized the dose statistics for this structure as a plan quality factor to determine the optimal treatment plan. All the CTV and OARs contours on CT and CBCT were created and reviewed by radiation oncologists to ensure the accuracy and precision for dosimetry analyses. All the CBCT images, used for treatment planning and plan evaluation, were corrected to match the CT numbers of planning CT acquired from the institutional Siemens scanner (Chang *et al.*, 2023b).

**Table 1.** Clinical parameters and dose constraints for two-fraction prostate SBRT treatment planning.

| Clinical parameters | 2STAR | 2SMART |
|---|---|---|
| Prescription | 26 Gy | 32 Gy |
| Fraction (Fx) | 2 | 2 |
| Beams | 4 | 4 |
| CTV | D98 ≥ 100% | D98 ≥ 100% |
|  | D0.03cc < 110% | D0.03cc < 110% |
| Bladder | V14.6Gy < 15 cc | V14.6Gy < 25 cc |
|  | V20.8Gy < 5 cc | V20.8Gy < 10 cc |
| Urethra | N/A | D0.03cc < 33.8 Gy |
|  |  | D10% < 30.4 Gy |
| Rectum | V13Gy < 7 cc | V13Gy < 7 cc |
|  | V17.6Gy < 4 cc | V17.6Gy < 4 cc |
|  | V20.8Gy < 1 cc | V22Gy < 1 cc |

### 2.3 Digital twin framework for adaptive proton therapy

The proposed DT framework aims to generate multiple treatment plans utilizing patient-specific CTV robust positional setups with associated uncertainty, facilitating optimal treatment plan delivery based on same-day CBCT evaluation. Figure 1 illustrates the proposed DT framework for adaptive proton therapy in a two-fraction prostate SBRT treatment scenario. We streamline the entire treatment planning process to place greater emphasis on comparing conventional clinical planning with DT-based planning. The initial pre-treatment planning (pCT) is utilized for generating the treatment plan for Fraction 1 (Fx1), denoted as Plan 1 (Element 1). In the clinical workflow, this plan undergoes robust optimization with 5 mm CTV positional uncertainty, except posteriorly, where a 3-mm margin is applied (Element 2), in alignment with the planning goals outlined in Table 1.

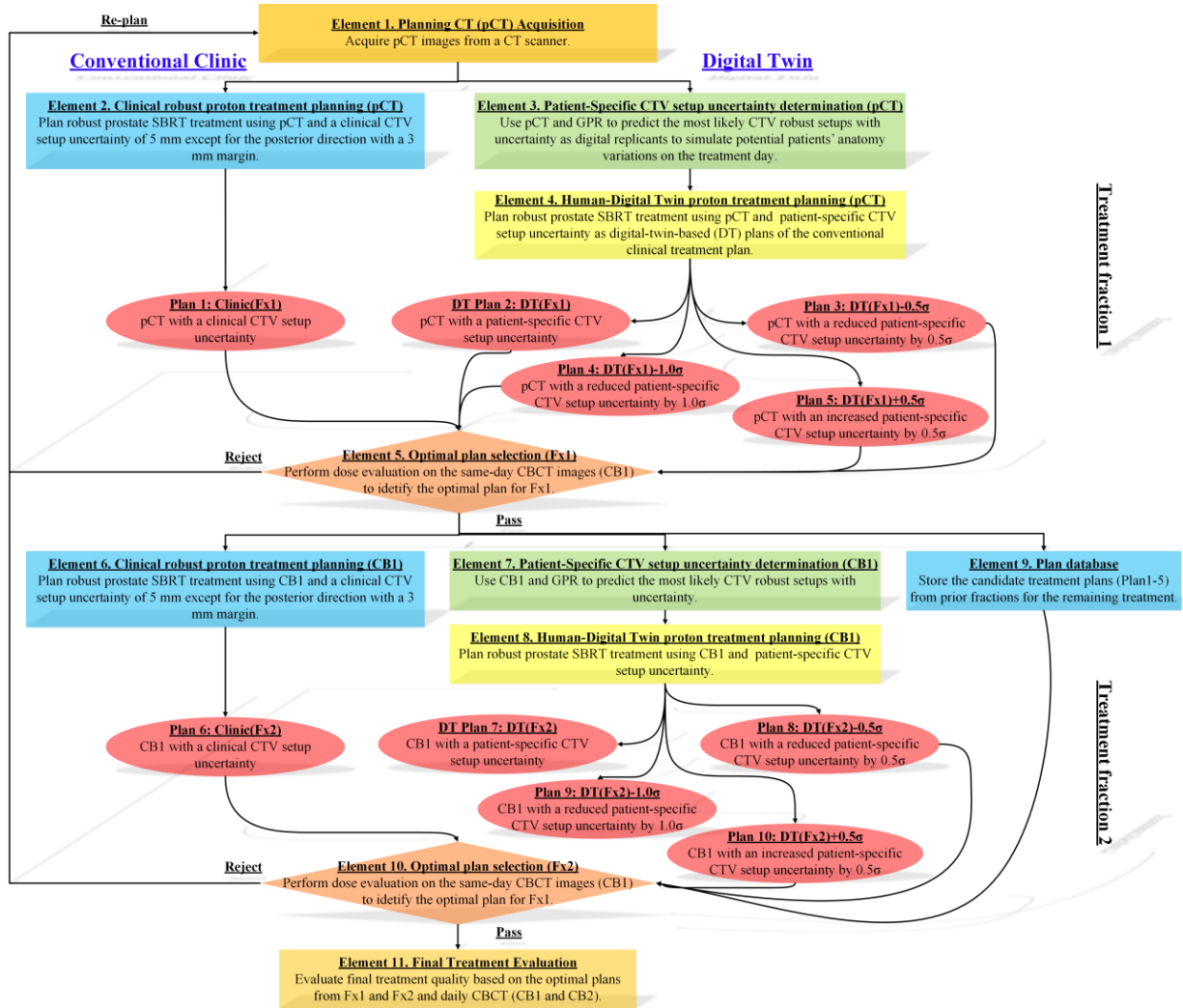

**Figure 1.** Overview of the proposed digital twin (DT) framework for adaptive two-fraction proton prostate SBRT treatment. The framework generates digital-twin-based treatment plans that enable the plan selection on the treatment day. Initially, pre-treatment images are acquired from a CT scanner for planning (pCT) (Element 1). The clinical robust proton plan is obtained with a standard CTV setup uncertainty for all patients (Element 2), and the clinical Plan 1 is also a part of candidate plans in the framework that should be considered while searching the optimal treatment plan. The DT-based plans (Plan 2-5) are robustly optimized based on patient-specific CTV setups with uncertainty (Element 3-4). The optimal treatment is determined by plan evaluation using the first CBCT (CB1) before treatment fraction 1 (Fx1) delivery (Element 5). Then CB1 will be used to create a clinical plan (Plan 6) and DT-based plans (Plan 7-10) as shown in Element 6-8. The second treatment fraction (Fx2) CBCT (CB2) is used to identify the optimal plan (Element 10) including prior Plan 1-5 (Element 9). Ultimately, the overall treatment quality will be evaluated (Element 11) using the optimal plans from Fx1 and Fx2 and daily CBCT images (CB1 and CB2). The σ denotes the uncertainty of CTV positional setups.

Element 3 in Figure 1 delineates another planning pathway employing DT. The patient-specific CTV setup uncertainty are predicted utilizing Gaussian Process Regression (GPR) (Ebden, 2015) based on pCT image features. Further details regarding the GPR models are provided in Section 2.4. The GPR models forecast the most probable CTV positional setup along with associated uncertainty, leading to the generation of four distinct robust positional uncertainty setups for plan optimization, including the most probable setup, ±0.5 standard deviation ($\sigma$) margins, and -$\sigma$ margin for Plans 2-5 (Element 4). These DT-based robust CTV margins are additionally constrained within upper and lower limits of 1.5 mm and 5.0 mm. The lower bound accounts for inherent uncertainty stemming from the alignment between radiation and mechanical isocenters, while the upper bound reflects institutional guidelines. All pre-Fx1 plans are optimized using pCT data but with varying robust CTV setup uncertainty.

On the day of fraction 1 treatment (Fx1), Plans 1-5 depicted in Figure 1 undergo evaluation using the fraction 1 CBCT (CB1) through our previously developed online CBCT evaluation framework (Chang *et al.*, 2023b) to select the optimal plan for Fx1 treatment delivery (Element 5). The optimal plan selection is based on factors such as CTV coverage and sparing of OARs, with detailed methods for quantifying plan quality provided in Section 2.5. In case no plan meets our clinical goals during CB1 evaluation, a re-plan request will be initiated, necessitating a restart of the workflow from Element 1.

Following Fx1, we update the planning images using CB1 with the clinical CTV setup uncertainty to create clinical Plan 6 (Element 6). Then CB1 images are used along with a new set of patient-specific CTV derived from CB1 to perform robust plan optimization to generate DT-based treatment plans, Plans 7-10 (Elements 7-8). Element 9 in Figure 1 illustrates that previous treatment plans for Fx1 become part of the plan database, which can be utilized to expand the candidate plans for optimal solution search. For fraction 2 treatment (Fx2), plan evaluation is based on the fraction 2 CBCT (CB2), incorporating candidate plans including Plan 1-10 from Figure 1 for optimal delivery plan search (Element 10). DT-based planning yields a total of 100 potential robust treatment plans with various patient-specific CTV setup uncertainties and manually optimized by medical physicists across 10 patients, necessitating the evaluation of 150 plans using CB1 and CB2 before optimal treatment delivery plans are identified. Ultimately, the optimal plans from Fx1 and Fx2 are jointly evaluated to examine the plan quality of the two-fraction prostate SBRT.

### 2.4 Gaussian process regression for predicting patient-specific CTV robust setups with uncertainty

We utilized a MATLAB-based GPR package (Rasmussen and Nickisch, 2010) to find the underlying correlation between the relative CTV position ($\Delta$CTV) and image features extracted from both CT and CBCT scans. The image features were manually defined using internal functions within RayStation 2023B through scripting. Initially, we computed the center of mass coordinates of CTV contours as the reference position of CTV. Subsequently, we constructed a bounding box encompassing the femoral heads, and delineated the reference femoral coordinates based on the vertices of this bounding box. The coordinate for the left femoral head (FemL) was defined as the vertex posterior and inferior to the patient's midline, while the coordinate for the right femoral head (FemR) was defined as the vertex posterior and superior to the patient's midline. Then $\Delta$CTV was defined by using CTV coordinate subtracting FemL coordinate in left-right, anterior-posterior, and superior-inferior directions. The $\Delta$CTV was then determined by subtracting the CTV coordinate from the FemL coordinate. Table 2 describes all the image features utilized to correlate with $\Delta$CTV.

**Table 2.** Definition of image features utilized for constructing a Gaussian Process Regression (GPR) model that correlates with patient-specific relative CTV position (ΔCTV).

| Feature | Description |
| --- | --- |
| 1 | Euclidean distance between FemL and FemR |
| 2 | Euclidean distance between CTV and FemL |
| 3 | Euclidean distance between CTV and FemR |
| 4 | Angle between the two vectors by CTV-FemL and CTV-FemR |
| 5 | Distance ($\Delta X$) between FemR and CTV in left-right direction |
| 6 | Distance ($\Delta Y$) between FemR and CTV in anterior-posterior direction |
| 7 | Distance ($\Delta Z$) between FemR and CTV in superior-inferior direction |

### 2.5 Treatment plan selection using ProKnow scoring system and same-day CBCT

The proposed DT framework generates multiple candidate plans to facilitate optimal plan delivery on the treatment day. Each plan evaluation relies on the CBCT images acquired on the treatment day to ensure the accuracy of patients' anatomy. To systematically assess plan quality, we adopted the ProKnow® (ProKnow Systems, Sanford, FL, USA)) (Nelms *et al.*, 2012), which is a scoring method previously utilized in the 2016 AAMD/RSS-SBRT Prostate (Richard Sweat *et al.*, 2016) and plan quality studies (Gao *et al.*, 2023), and adapted the system for the two-fraction prostate proton SBRT treatment regimen. The ProKnow system assigns varying scores based on cumulative doses from CTV and OARs, with higher scores indicating superior clinical dosimetry outcomes compared to lower-scored plans. Using this system, we identify the optimal plan as the one achieving the highest scores among all candidate plans for the same patient at the same treatment fraction. Since the scoring system includes dose conformality metrics, we further define a volume (*V*) as an expansion from the CTV by 5 mm in all directions except posteriorly, using a 3-mm expansion. The *V* is used only for calculating dose conformality as the evaluation metrics in the ProKnow scoring system. Figure 2 illustrates our scoring system consisting of 12 scoring functions, adapted from the original ProKnow scoring functions based on our institutional planning guidelines and the reported two-fraction prostate SBRT clinical trials (i.e., 2STAR and 2SMART). Each scoring function corresponds to one plan quality metric. The involved plan quality metrics are V100 (the percentage of relative volume receiving 100% CTV prescription dose), D98 (the percentage of relative dose received by 98% CTV volume), D2cm (maximum dose at 2 cm from *V* relative to the prescription dose) which represent the dose fall-off speed, Paddick Conformity Index (PCI) (Paddick, 2000), D0.03cc (dose received by the highest irradiated 0.03 cm$^3$) which represents the hot spot of the proton plan, the mean dose and V100 of the bladder neck, V20.8Gy and V14.6Gy of the bladder, as well as V20.8Gy, V17.6Gy, and V13Gy of the rectum. The vertical dashed lines in the figure denote the planning goals from Table 1.

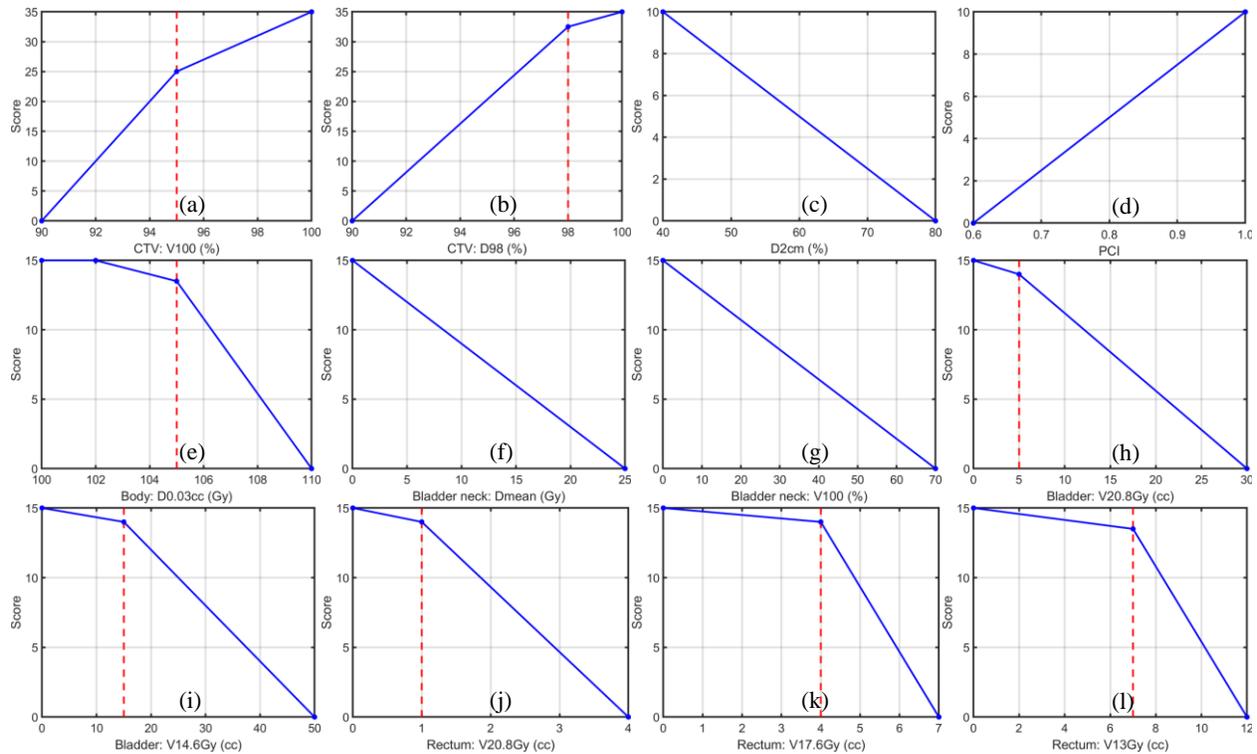

**Figure 2.** ProKnow scoring functions for quantifying the quality of clinical plans and DT-based plans. The scoring system is based on the cumulative doses from evaluated plans on the same-day CBCT, which shows the actual patients' anatomy on the treatment day. (a)-(l) shows the scoring functions for different dosimetric parameters. The vertical red dashed lines depict the clinical planning goals.

## 3 Results

### 3.1 Patient-specific CTV setup uncertainty

Figure 3 illustrates patient-specific robust CTV setup uncertainty in the left-right, anterior-posterior, and superior-inferior directions, which were predicted by the trained GPR model using the features extracted from pre-treatment pCT images as depicted in Figure 1 (Element 3). These setup uncertainties were employed by DT-based planning for the first treatment fraction. Error bars represent a range of ±0.5σ CTV uncertainty margins, bounded by 1.5 mm due to inherent uncertainties arising from the coincidence of radiation and mechanical isocenters. An additional 5 mm upper bound is imposed by our institutional clinical guidelines. Figure 4 displays the robust CTV setup uncertainty for all DT-based planning on CBCT image set CB1 utilized for the second treatment fraction.

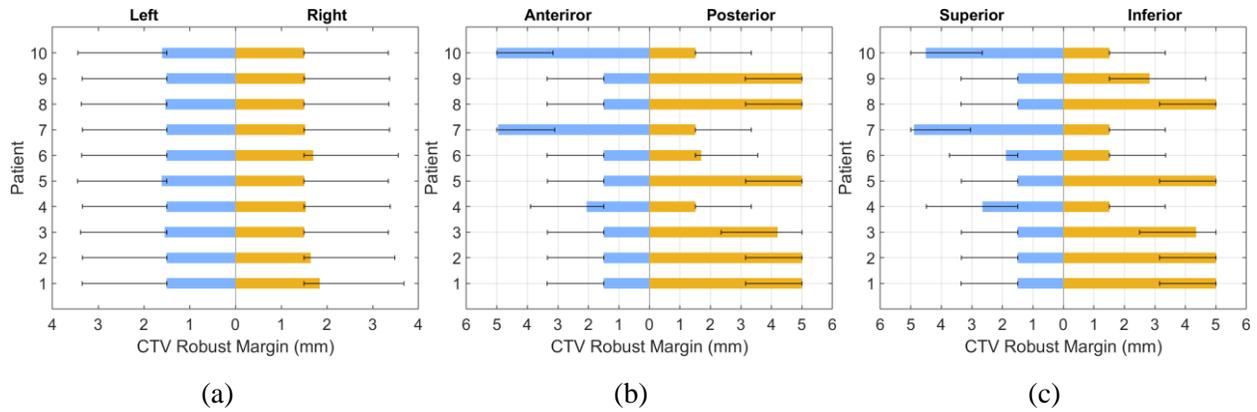

**Figure 3.** Patient-specific robust CTV setup uncertainty based on pCT for DT-based plans (treatment fraction 1) in (a) left-right, (b) anterior-posterior, and (c) superior-inferior directions. The error bars denote the uncertainty of ±0.5σ margins.

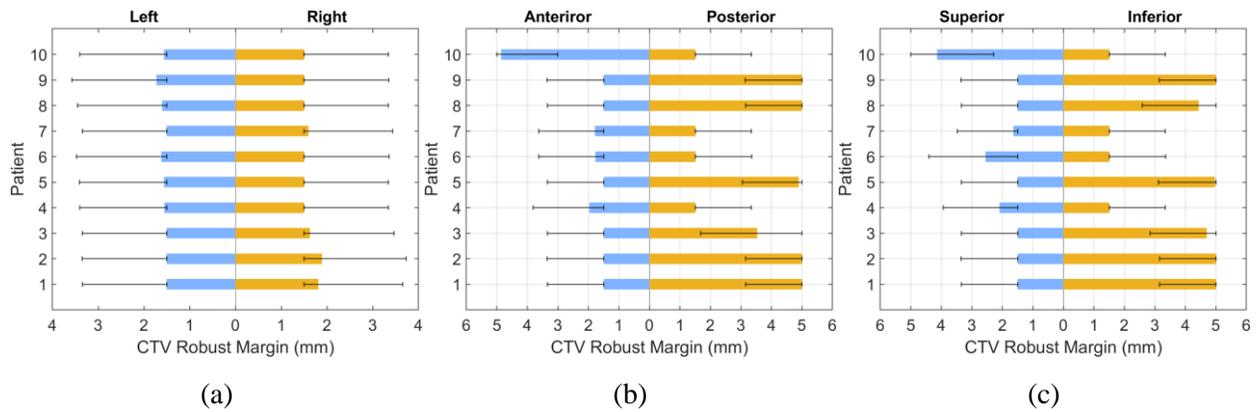

**Figure 4.** Patient-specific robust CTV setup uncertainty based on CB1 for DT-based plans (treatment fraction 2) in (a) left-right, (b) anterior-posterior, and (c) superior-inferior directions. The error bars denote the uncertainty of ±0.5σ margins.

### 3.2 DT-based plan selection for prostate SBRT fraction 1

For treatment Fx1, there are a total of 5 candidate plans including 1 clinical and 4 DT-based plans as depicted in Figure 1 (Element 5). Table 3 presents the dose evaluation results of clinical and DT-based treatment plans for Fx1 calculated using CB1. The plans were generated utilizing proton robust optimization with pCT. Overall, both types of plans exhibit similar CTV coverage, with minor discrepancies. Notably, DT-based planning, particularly for Patient 9, demonstrates an 11% increase in V100 coverage compared to clinical planning. Additionally, the majority of DT-based plans showcase superior sparing of OARs compared to clinical plans. For instance, in Patient 3, the DT-based plan reduces bladder neck V100 by 18.2% compared to the clinical plan. Figure 5 illustrates the obtained plan scores calculated using the modified ProKnow scoring system, which demonstrates that most DT-based plans receive higher scores than clinical plans.

**Table 3.** Dose volume endpoint comparisons of CTV and OAR for the ten prostate SBRT patients, obtained by the clinical plans (clinic) and the plans by the proposed digital twin (DT) framework based on pCT images. Dose statistics were derived from plan evaluation using CB1 images before delivering treatment fraction 1 (Fx1). The σ denotes the uncertainty of CTV robust optimization margins for DT-based treatment planning.

| pCT (planning) CB1 (Evaluation) | | CTV | | Conformality | | Bladder neck | | Body | Bladder | | Rectum | | |
|---|---|---|---|---|---|---|---|---|---|---|---|---|---|
| | | V100 (%) | D98 (%) | D2cm (%) | PCI | V100 (%) | Dmean (Gy) | D0.03cc (%) | V20.8Gy (cc) | V14.6Gy (cc) | V20.8Gy (cc) | V17.6Gy (cc) | V13Gy (cc) |
| P01 | Clinic(Fx1) | 98.07 | 100.03 | 79.28 | 0.75 | 14.49 | 18.36 | 107.11 | 4.80 | 10.60 | 1.05 | 1.81 | 3.80 |
| | DT(Fx1) | 97.73 | 99.73 | 72.30 | 0.75 | 10.58 | 18.02 | 106.03 | 3.12 | 8.52 | 1.14 | 1.93 | 3.77 |
| | DT(Fx1)-0.5σ | 97.90 | 99.69 | 73.58 | 0.76 | 12.36 | 18.03 | 106.37 | 3.13 | 8.58 | 1.15 | 1.97 | 3.82 |
| | DT(Fx1)-1.0σ | 97.90 | 99.89 | 64.81 | 0.76 | 12.50 | 18.08 | 106.71 | 3.00 | 8.35 | 1.11 | 1.92 | 3.55 |
| | DT(Fx1)+0.5σ | 97.80 | 99.83 | 77.42 | 0.75 | 12.40 | 18.27 | 105.78 | 3.96 | 9.57 | 1.08 | 1.92 | 3.88 |
| P02 | Clinic(Fx1) | 98.10 | 100.03 | 52.19 | 0.74 | 19.79 | 16.54 | 107.61 | 3.34 | 6.75 | 1.32 | 2.98 | 7.25 |
| | DT(Fx1) | 98.76 | 100.56 | 52.17 | 0.75 | 4.03 | 15.02 | 108.02 | 2.42 | 5.59 | 1.74 | 4.08 | 8.46 |
| | DT(Fx1)-0.5σ | 98.48 | 100.26 | 51.21 | 0.74 | 7.83 | 15.09 | 107.28 | 2.56 | 5.69 | 1.40 | 3.08 | 7.31 |
| | DT(Fx1)-1.0σ | 98.03 | 100.00 | 52.12 | 0.73 | 6.62 | 15.13 | 107.08 | 2.56 | 5.67 | 1.30 | 2.93 | 6.87 |
| | DT(Fx1)+0.5σ | 99.53 | 100.97 | 52.08 | 0.77 | 13.90 | 16.68 | 108.24 | 3.56 | 7.28 | 1.78 | 4.15 | 8.61 |
| P03 | Clinic(Fx1) | 98.83 | 100.37 | 50.08 | 0.82 | 25.84 | 21.99 | 108.45 | 8.00 | 15.52 | 0.68 | 2.29 | 5.30 |
| | DT(Fx1) | 98.11 | 100.09 | 46.76 | 0.78 | 7.64 | 19.38 | 107.68 | 5.11 | 11.87 | 0.68 | 2.14 | 5.04 |
| | DT(Fx1)-0.5σ | 98.12 | 100.05 | 47.25 | 0.78 | 5.72 | 19.25 | 108.30 | 5.12 | 11.82 | 0.69 | 2.22 | 5.11 |
| | DT(Fx1)-1.0σ | 98.33 | 100.23 | 46.77 | 0.79 | 8.15 | 19.31 | 107.99 | 5.11 | 11.84 | 0.69 | 2.17 | 5.10 |
| | DT(Fx1)+0.5σ | 98.37 | 100.25 | 46.03 | 0.81 | 10.99 | 20.36 | 108.07 | 6.24 | 13.12 | 0.71 | 2.25 | 5.19 |
| P04 | Clinic(Fx1) | 96.78 | 99.86 | 50.52 | 0.76 | 22.15 | 13.57 | 108.44 | 13.75 | 24.73 | 3.20 | 5.31 | 8.77 |
| | DT(Fx1) | 96.18 | 99.79 | 49.67 | 0.76 | 21.55 | 13.38 | 108.80 | 13.67 | 24.56 | 3.15 | 5.29 | 8.61 |
| | DT(Fx1)-0.5σ | 95.32 | 99.67 | 49.38 | 0.75 | 19.15 | 13.36 | 108.67 | 13.76 | 24.48 | 3.10 | 5.22 | 8.50 |
| | DT(Fx1)-1.0σ | 95.32 | 99.67 | 49.38 | 0.75 | 19.15 | 13.36 | 108.67 | 13.76 | 24.48 | 3.10 | 5.22 | 8.50 |
| | DT(Fx1)+0.5σ | 96.62 | 99.77 | 51.97 | 0.76 | 22.54 | 13.66 | 108.48 | 13.84 | 24.67 | 3.22 | 5.39 | 8.89 |
| P05 | Clinic(Fx1) | 98.67 | 100.28 | 53.22 | 0.80 | 18.85 | 18.15 | 106.44 | 5.01 | 11.21 | 3.86 | 6.35 | 10.93 |
| | DT(Fx1) | 98.82 | 100.17 | 51.23 | 0.79 | 11.29 | 17.34 | 105.91 | 4.13 | 10.09 | 3.86 | 6.42 | 10.88 |
| | DT(Fx1)-0.5σ | 98.62 | 100.17 | 49.12 | 0.79 | 11.32 | 17.39 | 105.72 | 4.13 | 10.18 | 4.00 | 6.41 | 10.89 |
| | DT(Fx1)-1.0σ | 98.35 | 100.17 | 48.45 | 0.79 | 11.29 | 17.35 | 105.68 | 4.11 | 10.14 | 3.98 | 6.45 | 10.51 |
| | DT(Fx1)+0.5σ | 98.81 | 100.22 | 51.56 | 0.80 | 13.36 | 17.52 | 106.17 | 4.28 | 10.39 | 3.88 | 6.40 | 10.88 |

(Continuous)

| pCT (planning) CB1 (Evaluation) | | CTV | | Conformality | | Bladder neck | | Body | Bladder | | Rectum | | |
|---|---|---|---|---|---|---|---|---|---|---|---|---|---|
| | | V100 (%) | D98 (%) | D2cm (%) | PCI | V100 (%) | Dmean (Gy) | D0.03cc (%) | V20.8Gy (cc) | V14.6Gy (cc) | V20.8Gy (cc) | V17.6Gy (cc) | V13Gy (cc) |
| P06 | Clinic(Fx1) | 99.91 | 100.93 | 64.42 | 0.82 | 64.03 | 25.16 | 107.80 | 15.49 | 26.53 | 0.22 | 0.87 | 2.79 |
| | DT(Fx1) | 99.21 | 100.48 | 63.23 | 0.80 | 24.94 | 22.75 | 106.56 | 8.96 | 18.39 | 0.21 | 0.57 | 2.21 |
| | DT(Fx1)-0.5σ | 98.70 | 100.23 | 58.90 | 0.79 | 19.96 | 22.41 | 106.50 | 8.63 | 17.72 | 0.19 | 0.52 | 2.09 |
| | DT(Fx1)-1.0σ | 98.70 | 100.23 | 58.90 | 0.79 | 19.96 | 22.41 | 106.50 | 8.63 | 17.72 | 0.19 | 0.52 | 2.09 |
| | DT(Fx1)+0.5σ | 99.30 | 100.59 | 64.38 | 0.81 | 24.39 | 24.39 | 107.23 | 12.08 | 22.04 | 0.21 | 0.61 | 2.50 |
| P07 | Clinic(Fx1) | 96.21 | 99.76 | 73.55 | 0.76 | 49.50 | 21.88 | 106.70 | 15.79 | 23.88 | 0.16 | 0.96 | 4.39 |
| | DT(Fx1) | 96.09 | 99.72 | 72.26 | 0.75 | 48.41 | 21.84 | 105.99 | 15.26 | 23.46 | 0.08 | 0.80 | 3.76 |
| | DT(Fx1)-0.5σ | 98.55 | 100.13 | 67.46 | 0.77 | 43.16 | 20.95 | 105.67 | 12.75 | 21.02 | 0.18 | 0.88 | 3.63 |
| | DT(Fx1)-1.0σ | 99.81 | 100.62 | 62.00 | 0.79 | 36.32 | 20.45 | 105.54 | 11.55 | 19.60 | 0.09 | 0.72 | 3.35 |
| | DT(Fx1)+0.5σ | 97.17 | 99.76 | 73.37 | 0.77 | 50.80 | 21.74 | 106.65 | 15.41 | 23.92 | 0.25 | 0.91 | 4.12 |
| P08 | Clinic(Fx1) | 99.16 | 101.16 | 74.92 | 0.77 | 54.95 | 25.13 | 108.38 | 16.39 | 31.75 | 2.18 | 3.62 | 6.28 |
| | DT(Fx1) | 99.24 | 101.67 | 73.25 | 0.78 | 18.27 | 24.19 | 107.45 | 14.55 | 28.93 | 3.06 | 4.80 | 8.46 |
| | DT(Fx1)-0.5σ | 99.38 | 101.13 | 69.46 | 0.79 | 18.11 | 24.25 | 107.42 | 14.71 | 29.07 | 2.29 | 3.79 | 6.72 |
| | DT(Fx1)-1.0σ | 99.28 | 100.71 | 63.24 | 0.80 | 25.87 | 24.30 | 107.22 | 14.67 | 27.56 | 2.09 | 3.70 | 6.31 |
| | DT(Fx1)+0.5σ | 99.56 | 100.99 | 76.73 | 0.77 | 33.76 | 24.61 | 107.65 | 15.53 | 31.24 | 2.81 | 4.35 | 8.04 |
| P09 | Clinic(Fx1) | 87.59 | 99.05 | 57.67 | 0.69 | 62.57 | 24.57 | 104.95 | 28.45 | 43.57 | 2.43 | 3.17 | 5.70 |
| | DT(Fx1) | 98.59 | 100.12 | 55.82 | 0.71 | 61.50 | 24.88 | 105.33 | 26.35 | 41.98 | 2.60 | 3.52 | 5.81 |
| | DT(Fx1)-0.5σ | 94.33 | 99.67 | 56.44 | 0.71 | 60.59 | 24.43 | 105.34 | 26.94 | 42.57 | 2.37 | 3.08 | 5.66 |
| | DT(Fx1)-1.0σ | 95.86 | 99.56 | 54.48 | 0.70 | 63.96 | 24.57 | 104.79 | 27.82 | 42.36 | 1.92 | 2.65 | 4.96 |
| | DT(Fx1)+0.5σ | 96.04 | 99.84 | 57.03 | 0.71 | 61.20 | 24.78 | 105.52 | 25.94 | 42.47 | 2.69 | 3.62 | 6.03 |
| P10 | Clinic(Fx1) | 98.94 | 100.37 | 57.68 | 0.82 | 29.47 | 14.67 | 108.02 | 24.72 | 39.77 | 1.55 | 2.84 | 5.29 |
| | DT(Fx1) | 99.09 | 100.41 | 54.02 | 0.83 | 23.06 | 13.80 | 107.49 | 22.59 | 36.83 | 1.04 | 2.09 | 4.38 |
| | DT(Fx1)-0.5σ | 99.26 | 100.58 | 45.83 | 0.82 | 22.33 | 13.55 | 106.89 | 20.58 | 35.21 | 1.03 | 2.09 | 4.34 |
| | DT(Fx1)-1.0σ | 99.22 | 100.52 | 46.94 | 0.83 | 22.74 | 13.50 | 108.34 | 20.35 | 34.68 | 1.07 | 2.11 | 4.37 |
| | DT(Fx1)+0.5σ | 99.04 | 100.55 | 55.60 | 0.84 | 27.02 | 14.26 | 107.71 | 23.83 | 38.27 | 1.15 | 2.20 | 4.48 |

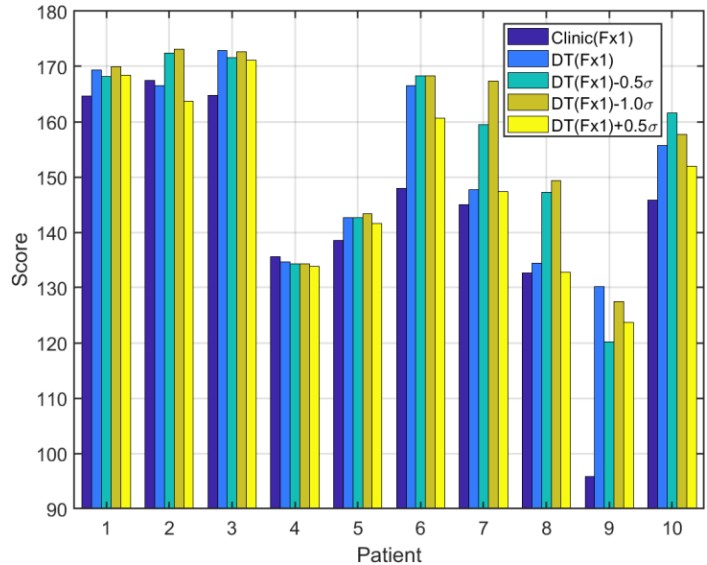

**Figure 5.** ProKnow score comparisons between the clinical plan and DT-based plans with different patient-specific CTV setup uncertainty. The plans were optimized on pCT, and the scores were derived based on the dosimetry parameters evaluated on CB1 from Table 3 and scoring functions in Figure 2.

### 3.3 DT-based plan selection for prostate SBRT fraction 2

For Fx2 treatment, there are a total of 10 candidate plans, including 5 plans carried over from Fx1, as illustrated in Figure 1 (Element 10). Table 4 compares the doses of the 5 plans carried over from Fx1 (i.e., the clinical and 4 DT-based treatment plans which were generated employing proton robust optimization with pCT) recalculated on CB2 for Fx2 evaluation. Similarly, Table 5 assesses the actual doses of the other five treatment plans, which were optimized using CB1, recalculated on CB2. For each patient, the optimal treatment delivery plan is determined based on the obtained dosimetry endpoints listed in both tables. Figures 6(a)-(b) display the plan scores corresponding to each plan listed in Table 4 and Table 5, respectively. Notably, Figure 6 reveals that 50% of the optimal plans originate from pCT-based plans, while the remaining 50% stem from updated planning images, CB1.

**Table 4.** Dose volume endpoint comparisons of CTV and OAR for the ten prostate SBRT patients, obtained by the clinical plans (clinic) and the plans by the proposed digital twin (DT) framework based on pCT images. Dose statistics were derived from plan evaluation using CB2 images before delivering treatment fraction 2 (Fx2). The σ denotes the uncertainty of CTV robust optimization margins for DT-based treatment planning.

| pCT (planning) CB2 (Evaluation) | | CTV | | Conformality | | Bladder neck | | Body | Bladder | | Rectum | | |
|---|---|---|---|---|---|---|---|---|---|---|---|---|---|
| | | V100 (%) | D98 (%) | D2cm (%) | PCI | V100 (%) | Dmean (Gy) | D0.03cc (%) | V20.8Gy (cc) | V14.6Gy (cc) | V20.8Gy (cc) | V17.6Gy (cc) | V13Gy (cc) |
| P01 | Clinic(Fx2) | 98.30 | 100.42 | 71.84 | 0.75 | 11.42 | 17.23 | 105.63 | 4.39 | 10.17 | 1.04 | 2.70 | 5.62 |
| | DT(Fx2) | 97.73 | 99.79 | 66.21 | 0.75 | 9.76 | 16.81 | 105.92 | 2.98 | 7.82 | 1.08 | 2.55 | 5.63 |
| | DT(Fx2)-0.5σ | 97.85 | 99.60 | 64.22 | 0.76 | 8.60 | 16.82 | 105.88 | 3.00 | 7.92 | 1.04 | 2.56 | 5.20 |
| | DT(Fx2)-1.0σ | 97.90 | 99.90 | 60.60 | 0.77 | 8.60 | 16.91 | 105.36 | 3.11 | 7.98 | 1.04 | 2.49 | 4.98 |
| | DT(Fx2)+0.5σ | 98.16 | 100.11 | 64.24 | 0.77 | 11.93 | 17.17 | 106.36 | 3.76 | 8.93 | 1.02 | 2.63 | 5.64 |
| P02 | Clinic(Fx2) | 98.27 | 100.22 | 53.40 | 0.77 | 24.98 | 22.27 | 107.50 | 4.29 | 9.92 | 1.88 | 3.16 | 5.39 |
| | DT(Fx2) | 97.95 | 99.94 | 52.35 | 0.77 | 6.36 | 20.99 | 106.80 | 2.96 | 8.21 | 2.29 | 3.63 | 5.87 |
| | DT(Fx2)-0.5σ | 98.07 | 100.17 | 51.85 | 0.77 | 8.47 | 21.01 | 106.71 | 3.05 | 8.27 | 2.00 | 3.27 | 5.46 |
| | DT(Fx2)-1.0σ | 98.06 | 100.12 | 52.55 | 0.76 | 7.64 | 21.06 | 106.68 | 3.10 | 8.37 | 1.92 | 3.18 | 5.43 |
| | DT(Fx2)+0.5σ | 99.00 | 101.00 | 50.53 | 0.79 | 20.19 | 22.51 | 107.55 | 4.71 | 10.67 | 2.26 | 3.72 | 5.98 |
| P03 | Clinic(Fx2) | 98.00 | 100.00 | 50.06 | 0.76 | 21.86 | 22.55 | 109.68 | 7.81 | 15.87 | 0.92 | 2.28 | 5.23 |
| | DT(Fx2) | 96.43 | 98.86 | 47.09 | 0.69 | 6.18 | 20.09 | 108.97 | 4.89 | 12.01 | 0.85 | 2.22 | 4.96 |
| | DT(Fx2)-0.5σ | 96.72 | 98.87 | 47.73 | 0.70 | 4.73 | 19.93 | 108.85 | 4.81 | 11.98 | 0.84 | 2.21 | 4.97 |
| | DT(Fx2)-1.0σ | 96.95 | 98.92 | 47.13 | 0.70 | 5.68 | 20.05 | 109.30 | 4.84 | 12.04 | 0.86 | 2.23 | 5.04 |
| | DT(Fx2)+0.5σ | 97.48 | 99.62 | 46.22 | 0.73 | 10.53 | 21.10 | 108.91 | 6.05 | 13.29 | 0.82 | 2.10 | 5.01 |
| P04 | Clinic(Fx2) | 94.36 | 99.58 | 53.11 | 0.80 | 35.64 | 15.80 | 108.91 | 13.48 | 23.34 | 0.05 | 0.22 | 1.10 |
| | DT(Fx2) | 94.15 | 99.51 | 51.28 | 0.80 | 34.66 | 15.62 | 108.69 | 13.22 | 23.22 | 0.03 | 0.19 | 1.05 |
| | DT(Fx2)-0.5σ | 93.70 | 99.56 | 51.11 | 0.80 | 34.28 | 15.62 | 108.33 | 13.17 | 23.15 | 0.04 | 0.18 | 1.03 |
| | DT(Fx2)-1.0σ | 93.70 | 99.56 | 51.11 | 0.80 | 34.28 | 15.62 | 108.33 | 13.17 | 23.15 | 0.04 | 0.18 | 1.03 |
| | DT(Fx2)+0.5σ | 94.77 | 99.60 | 53.04 | 0.81 | 36.03 | 15.90 | 108.67 | 13.63 | 23.45 | 0.04 | 0.19 | 1.12 |
| P05 | Clinic(Fx2) | 99.07 | 100.33 | 50.82 | 0.80 | 27.81 | 19.27 | 105.84 | 5.64 | 11.98 | 0.77 | 1.81 | 3.73 |
| | DT(Fx2) | 99.25 | 100.35 | 49.39 | 0.79 | 18.47 | 18.38 | 105.62 | 5.03 | 10.99 | 0.83 | 1.77 | 3.83 |
| | DT(Fx2)-0.5σ | 98.98 | 100.26 | 48.38 | 0.79 | 16.34 | 18.43 | 105.73 | 5.04 | 11.07 | 0.86 | 1.85 | 3.86 |
| | DT(Fx2)-1.0σ | 98.71 | 100.35 | 47.89 | 0.79 | 17.67 | 18.43 | 105.35 | 4.99 | 11.14 | 0.88 | 1.92 | 3.90 |
| | DT(Fx2)+0.5σ | 99.24 | 100.33 | 49.71 | 0.80 | 21.51 | 18.60 | 105.55 | 5.27 | 11.38 | 0.84 | 1.77 | 3.78 |

(Continuous)

| pCT (planning) CB2 (Evaluation) | | CTV | | Conformality | | Bladder neck | | Body | Bladder | | Rectum | | |
|---|---|---|---|---|---|---|---|---|---|---|---|---|---|
| | | V100 (%) | D98 (%) | D2cm (%) | PCI | V100 (%) | Dmean (Gy) | D0.03cc (%) | V20.8Gy (cc) | V14.6Gy (cc) | V20.8Gy (cc) | V17.6Gy (cc) | V13Gy (cc) |
| P06 | Clinic(Fx2) | 99.94 | 100.92 | 64.39 | 0.82 | 50.97 | 24.38 | 107.53 | 13.73 | 24.55 | 0.63 | 1.69 | 4.27 |
| | DT(Fx2) | 98.82 | 100.31 | 57.47 | 0.80 | 16.93 | 21.78 | 106.87 | 7.59 | 16.87 | 0.57 | 1.55 | 3.75 |
| | DT(Fx2)-0.5σ | 98.91 | 100.38 | 51.94 | 0.79 | 16.15 | 21.38 | 106.87 | 7.17 | 15.96 | 0.52 | 1.53 | 3.61 |
| | DT(Fx2)-1.0σ | 98.91 | 100.38 | 51.94 | 0.79 | 16.15 | 21.38 | 106.87 | 7.17 | 15.96 | 0.52 | 1.53 | 3.61 |
| | DT(Fx2)+0.5σ | 99.33 | 100.56 | 57.52 | 0.83 | 43.84 | 23.53 | 107.60 | 10.46 | 20.68 | 0.58 | 1.63 | 3.88 |
| P07 | Clinic(Fx2) | 96.50 | 99.73 | 76.91 | 0.79 | 34.38 | 17.88 | 106.72 | 16.21 | 25.70 | 1.68 | 4.32 | 9.15 |
| | DT(Fx2) | 97.12 | 99.73 | 75.61 | 0.78 | 30.32 | 17.91 | 105.72 | 16.16 | 25.02 | 1.35 | 3.95 | 8.46 |
| | DT(Fx2)-0.5σ | 98.24 | 100.03 | 71.89 | 0.79 | 16.74 | 16.74 | 105.85 | 13.44 | 22.43 | 1.35 | 4.23 | 8.60 |
| | DT(Fx2)-1.0σ | 99.65 | 100.66 | 69.01 | 0.80 | 12.82 | 16.01 | 106.09 | 11.71 | 20.66 | 1.31 | 3.71 | 7.89 |
| | DT(Fx2)+0.5σ | 97.42 | 99.92 | 76.59 | 0.80 | 30.81 | 17.86 | 106.40 | 15.94 | 25.31 | 1.60 | 4.34 | 8.93 |
| P08 | Clinic(Fx2) | 98.74 | 100.25 | 73.98 | 0.76 | 31.66 | 24.93 | 107.66 | 8.57 | 19.31 | 2.28 | 3.64 | 6.90 |
| | DT(Fx2) | 98.92 | 100.42 | 73.48 | 0.73 | 30.93 | 24.24 | 105.66 | 7.17 | 17.01 | 3.08 | 4.97 | 9.54 |
| | DT(Fx2)-0.5σ | 99.22 | 100.65 | 70.89 | 0.75 | 27.05 | 24.24 | 105.37 | 7.23 | 16.97 | 2.19 | 4.12 | 7.70 |
| | DT(Fx2)-1.0σ | 99.48 | 100.83 | 63.56 | 0.77 | 22.24 | 24.17 | 105.89 | 7.23 | 16.01 | 2.02 | 3.59 | 6.53 |
| | DT(Fx2)+0.5σ | 99.48 | 101.25 | 73.49 | 0.75 | 30.10 | 24.57 | 107.23 | 7.65 | 18.23 | 2.83 | 4.70 | 9.04 |
| P09 | Clinic(Fx2) | 100.00 | 101.75 | 58.63 | 0.77 | 55.08 | 23.70 | 104.51 | 25.04 | 39.36 | 0.70 | 1.05 | 3.03 |
| | DT(Fx2) | 100.00 | 101.81 | 56.17 | 0.79 | 53.93 | 23.86 | 105.01 | 23.25 | 37.79 | 0.51 | 1.54 | 2.93 |
| | DT(Fx2)-0.5σ | 100.00 | 101.66 | 56.50 | 0.78 | 50.71 | 23.32 | 104.41 | 23.26 | 37.73 | 0.47 | 1.30 | 2.77 |
| | DT(Fx2)-1.0σ | 99.90 | 101.53 | 53.16 | 0.78 | 53.34 | 23.33 | 104.94 | 23.89 | 38.20 | 0.39 | 0.71 | 2.53 |
| | DT(Fx2)+0.5σ | 100.00 | 101.81 | 57.19 | 0.78 | 53.10 | 23.73 | 104.36 | 23.36 | 37.64 | 0.51 | 1.53 | 3.00 |
| P10 | Clinic(Fx2) | 99.24 | 100.66 | 52.38 | 0.84 | 9.01 | 10.16 | 109.43 | 14.24 | 26.22 | 2.11 | 3.96 | 7.59 |
| | DT(Fx2) | 99.33 | 100.80 | 49.75 | 0.83 | 8.36 | 9.16 | 107.42 | 12.32 | 23.37 | 1.22 | 2.83 | 6.48 |
| | DT(Fx2)-0.5σ | 99.37 | 100.93 | 46.40 | 0.82 | 8.61 | 9.01 | 107.46 | 11.08 | 21.75 | 1.27 | 2.83 | 6.29 |
| | DT(Fx2)-1.0σ | 99.21 | 100.58 | 46.52 | 0.82 | 7.77 | 8.95 | 107.68 | 10.97 | 21.50 | 1.29 | 2.84 | 6.32 |
| | DT(Fx2)+0.5σ | 99.28 | 100.73 | 51.72 | 0.85 | 8.72 | 9.55 | 107.45 | 13.09 | 25.15 | 1.31 | 2.93 | 6.63 |

**Table 5.** Dose volume endpoint comparisons of CTV and OAR for the ten prostate SBRT patients, obtained by the clinical plans (clinic) and the plans by the proposed digital twin (DT) framework based on CB1 images. Dose statistics were derived from plan evaluation using CB2 images before delivering treatment fraction 2 (Fx2). The σ denotes the uncertainty of CTV robust optimization margins for DT-based treatment planning.

| CB1 (planning) CB2 (Evaluation) | | CTV | | Conformality | | Bladder neck | | Body | Bladder | | Rectum | | |
|---|---|---|---|---|---|---|---|---|---|---|---|---|---|
| | | V100 (%) | D98 (%) | D2cm (%) | PCI | V100 (%) | Dmean (Gy) | D0.03cc (%) | V20.8Gy (cc) | V14.6Gy (cc) | V20.8Gy (cc) | V17.6Gy (cc) | V13Gy (cc) |
| P01 | Clinic(Fx2) | 98.39 | 100.53 | 68.94 | 0.75 | 12.56 | 16.39 | 106.88 | 4.04 | 9.48 | 1.03 | 2.72 | 5.78 |
| | DT(Fx2) | 97.74 | 99.49 | 61.14 | 0.74 | 3.84 | 15.28 | 105.85 | 2.68 | 7.57 | 1.16 | 2.88 | 6.05 |
| | DT(Fx2)-0.5σ | 97.71 | 99.68 | 56.72 | 0.75 | 5.71 | 15.28 | 105.27 | 2.70 | 7.73 | 0.89 | 2.41 | 5.27 |
| | DT(Fx2)-1.0σ | 97.72 | 99.64 | 54.13 | 0.75 | 5.34 | 15.16 | 105.99 | 2.82 | 7.70 | 0.77 | 2.02 | 4.75 |
| | DT(Fx2)+0.5σ | 98.35 | 100.26 | 68.29 | 0.76 | 11.29 | 15.96 | 106.08 | 3.56 | 8.74 | 0.85 | 2.72 | 5.83 |
| P02 | Clinic(Fx2) | 98.82 | 101.05 | 49.33 | 0.80 | 51.75 | 25.31 | 106.82 | 8.21 | 15.53 | 1.92 | 3.09 | 4.96 |
| | DT(Fx2) | 96.68 | 98.75 | 52.34 | 0.75 | 34.27 | 24.59 | 106.32 | 6.15 | 12.62 | 1.80 | 2.98 | 5.02 |
| | DT(Fx2)-0.5σ | 96.90 | 98.84 | 48.74 | 0.76 | 29.48 | 24.44 | 106.93 | 6.23 | 12.79 | 1.74 | 2.82 | 4.55 |
| | DT(Fx2)-1.0σ | 97.73 | 99.78 | 48.96 | 0.77 | 29.32 | 24.43 | 106.73 | 6.34 | 13.11 | 1.69 | 2.63 | 4.38 |
| | DT(Fx2)+0.5σ | 96.43 | 98.57 | 49.44 | 0.74 | 30.80 | 24.53 | 105.88 | 6.31 | 13.49 | 1.83 | 3.00 | 4.98 |
| P03 | Clinic(Fx2) | 97.65 | 99.78 | 54.10 | 0.75 | 21.76 | 22.94 | 107.57 | 8.11 | 16.44 | 0.88 | 2.09 | 4.60 |
| | DT(Fx2) | 97.23 | 99.20 | 48.40 | 0.75 | 5.41 | 20.61 | 106.83 | 5.58 | 13.21 | 0.91 | 1.95 | 4.12 |
| | DT(Fx2)-0.5σ | 97.45 | 99.11 | 48.20 | 0.75 | 5.66 | 20.66 | 106.94 | 5.57 | 13.55 | 0.88 | 1.97 | 4.17 |
| | DT(Fx2)-1.0σ | 97.34 | 99.17 | 48.00 | 0.75 | 6.52 | 20.63 | 107.20 | 5.59 | 13.35 | 0.88 | 1.86 | 4.01 |
| | DT(Fx2)+0.5σ | 97.77 | 99.85 | 47.24 | 0.76 | 9.09 | 21.52 | 107.23 | 6.83 | 15.09 | 0.87 | 2.01 | 4.24 |
| P04 | Clinic(Fx2) | 94.80 | 99.64 | 54.59 | 0.82 | 28.88 | 14.80 | 108.92 | 14.22 | 23.85 | 0.02 | 0.11 | 0.55 |
| | DT(Fx2) | 94.76 | 99.60 | 50.98 | 0.82 | 31.94 | 14.80 | 108.50 | 14.28 | 23.72 | 0.02 | 0.11 | 0.51 |
| | DT(Fx2)-0.5σ | 96.17 | 99.73 | 48.13 | 0.82 | 34.17 | 14.89 | 108.92 | 14.54 | 24.22 | 0.01 | 0.04 | 0.30 |
| | DT(Fx2)-1.0σ | 96.17 | 99.73 | 48.13 | 0.82 | 34.17 | 14.89 | 108.92 | 14.54 | 24.22 | 0.01 | 0.04 | 0.30 |
| | DT(Fx2)+0.5σ | 95.38 | 99.62 | 47.69 | 0.82 | 28.10 | 15.07 | 107.67 | 16.31 | 27.89 | 0.00 | 0.00 | 0.12 |
| P05 | Clinic(Fx2) | 97.88 | 99.97 | 50.99 | 0.80 | 27.56 | 19.17 | 106.08 | 6.30 | 12.95 | 0.94 | 2.17 | 4.24 |
| | DT(Fx2) | 96.85 | 99.80 | 47.25 | 0.79 | 14.53 | 18.40 | 105.78 | 5.66 | 12.03 | 0.12 | 0.51 | 1.70 |
| | DT(Fx2)-0.5σ | 97.87 | 99.93 | 47.80 | 0.80 | 15.66 | 18.33 | 105.99 | 5.60 | 12.20 | 0.04 | 0.36 | 1.57 |
| | DT(Fx2)-1.0σ | 98.24 | 100.10 | 46.81 | 0.79 | 7.21 | 17.90 | 106.38 | 5.23 | 11.72 | 0.00 | 0.10 | 0.94 |
| | DT(Fx2)+0.5σ | 96.73 | 99.76 | 49.75 | 0.79 | 9.41 | 18.03 | 105.97 | 5.30 | 12.02 | 0.01 | 0.21 | 1.12 |

| CB1 (planning) CB2 (Evaluation) | | CTV | | Conformality | | Bladder neck | | Body | Bladder | | Rectum | | |
|---|---|---|---|---|---|---|---|---|---|---|---|---|---|
| | | V100 (%) | D98 (%) | D2cm (%) | PCI | V100 (%) | Dmean (Gy) | D0.03cc (%) | V20.8Gy (cc) | V14.6Gy (cc) | V20.8Gy (cc) | V17.6Gy (cc) | V13Gy (cc) |
| P06 | Clinic(Fx2) | 99.54 | 100.72 | 70.64 | 0.82 | 59.61 | 24.33 | 106.97 | 14.00 | 25.11 | 0.99 | 2.32 | 5.17 |
| | DT(Fx2) | 99.37 | 100.72 | 63.82 | 0.80 | 32.94 | 22.62 | 108.07 | 8.74 | 18.01 | 0.86 | 1.88 | 4.33 |
| | DT(Fx2)-0.5σ | 98.92 | 100.37 | 60.35 | 0.79 | 22.65 | 21.02 | 107.60 | 7.17 | 15.80 | 0.62 | 1.63 | 3.80 |
| | DT(Fx2)-1.0σ | 98.92 | 100.37 | 60.35 | 0.79 | 22.65 | 21.02 | 107.60 | 7.17 | 15.80 | 0.62 | 1.63 | 3.80 |
| | DT(Fx2)+0.5σ | 99.61 | 100.68 | 70.40 | 0.82 | 54.80 | 24.09 | 106.96 | 12.02 | 22.23 | 1.12 | 2.60 | 5.15 |
| P07 | Clinic(Fx2) | 99.96 | 100.83 | 73.04 | 0.79 | 35.42 | 18.58 | 105.61 | 17.35 | 27.02 | 2.45 | 5.10 | 8.82 |
| | DT(Fx2) | 99.64 | 100.53 | 67.58 | 0.80 | 13.94 | 16.27 | 105.19 | 10.62 | 19.40 | 0.62 | 2.06 | 4.91 |
| | DT(Fx2)-0.5σ | 99.73 | 101.10 | 65.46 | 0.80 | 13.57 | 15.78 | 105.21 | 10.96 | 19.57 | 0.93 | 2.26 | 5.37 |
| | DT(Fx2)-1.0σ | 99.73 | 101.10 | 65.46 | 0.80 | 13.57 | 15.78 | 105.21 | 10.96 | 19.57 | 0.93 | 2.26 | 5.37 |
| | DT(Fx2)+0.5σ | 99.07 | 100.32 | 70.36 | 0.80 | 21.37 | 17.41 | 105.23 | 12.98 | 21.69 | 1.67 | 4.05 | 7.37 |
| P08 | Clinic(Fx2) | 93.87 | 98.52 | 75.71 | 0.74 | 39.36 | 23.94 | 107.32 | 5.63 | 12.49 | 1.01 | 1.97 | 4.15 |
| | DT(Fx2) | 84.57 | 94.50 | 76.61 | 0.72 | 26.32 | 22.20 | 105.82 | 4.00 | 10.45 | 1.07 | 2.13 | 4.54 |
| | DT(Fx2)-0.5σ | 83.09 | 96.16 | 72.18 | 0.75 | 22.60 | 21.80 | 106.29 | 3.83 | 9.62 | 0.94 | 1.89 | 4.02 |
| | DT(Fx2)-1.0σ | 81.33 | 95.96 | 64.59 | 0.75 | 26.32 | 21.94 | 105.11 | 3.82 | 9.72 | 0.89 | 1.87 | 3.95 |
| | DT(Fx2)+0.5σ | 93.61 | 97.13 | 80.80 | 0.73 | 26.33 | 22.34 | 105.62 | 4.14 | 10.03 | 0.91 | 2.01 | 4.17 |
| P09 | Clinic(Fx2) | 67.38 | 93.88 | 62.15 | 0.73 | 21.52 | 18.14 | 106.59 | 15.01 | 27.49 | 2.37 | 4.34 | 8.27 |
| | DT(Fx2) | 69.47 | 93.25 | 65.23 | 0.71 | 16.75 | 17.55 | 106.08 | 10.72 | 22.06 | 2.64 | 4.39 | 8.08 |
| | DT(Fx2)-0.5σ | 67.69 | 93.08 | 64.82 | 0.72 | 13.03 | 17.45 | 106.08 | 11.64 | 24.33 | 2.08 | 3.60 | 7.57 |
| | DT(Fx2)-1.0σ | 72.64 | 93.17 | 61.78 | 0.74 | 18.65 | 16.41 | 105.81 | 12.04 | 22.86 | 2.14 | 3.91 | 7.51 |
| | DT(Fx2)+0.5σ | 63.13 | 92.09 | 70.69 | 0.69 | 12.92 | 17.31 | 107.25 | 10.71 | 23.67 | 2.85 | 4.89 | 8.62 |
| P10 | Clinic(Fx2) | 98.32 | 100.10 | 54.13 | 0.83 | 8.84 | 9.06 | 109.65 | 12.50 | 24.72 | 2.56 | 5.20 | 10.13 |
| | DT(Fx2) | 98.79 | 100.40 | 51.35 | 0.80 | 5.08 | 7.80 | 108.11 | 10.34 | 19.91 | 2.34 | 4.61 | 9.27 |
| | DT(Fx2)-0.5σ | 98.57 | 100.29 | 49.09 | 0.76 | 3.68 | 7.04 | 108.05 | 7.91 | 16.76 | 2.46 | 4.71 | 9.25 |
| | DT(Fx2)-1.0σ | 98.68 | 100.43 | 48.10 | 0.74 | 3.42 | 6.70 | 107.87 | 7.69 | 15.34 | 2.39 | 4.62 | 9.02 |
| | DT(Fx2)+0.5σ | 99.16 | 100.63 | 55.50 | 0.83 | 8.59 | 9.05 | 108.21 | 12.85 | 25.02 | 2.43 | 4.82 | 9.80 |

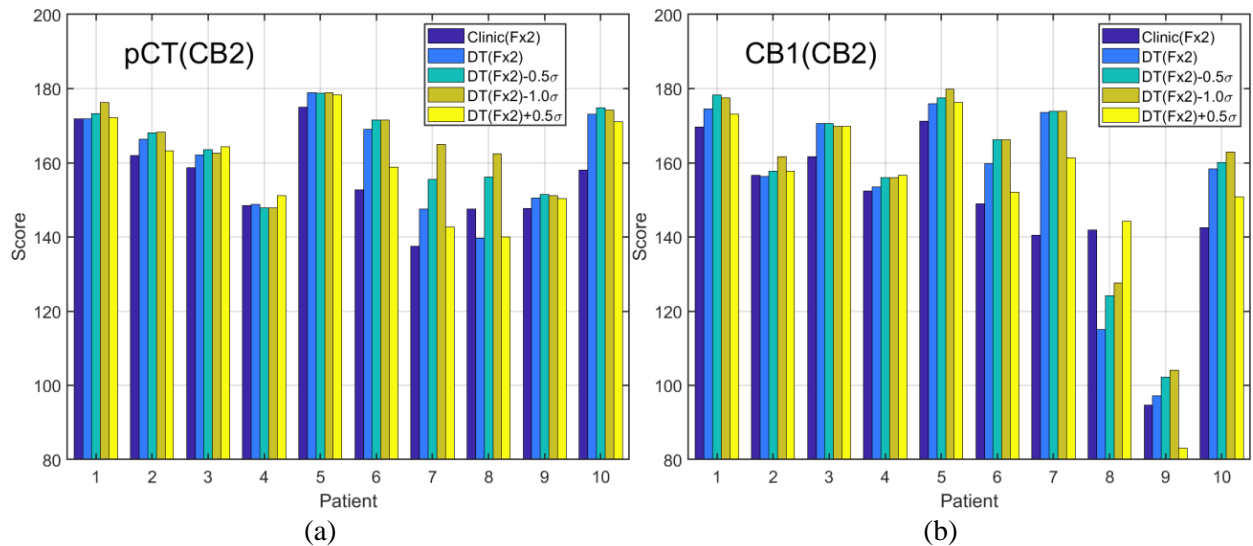

**Figure 6.** ProKnow score comparisons between the clinical plan and DT-based plans with different CTV setup uncertainty. (a) The plans were optimized on pCT, and the scores were derived based on the dosimetry parameters evaluated on CB2 from Table 4 and scoring functions in Figure 2. (b) The plans were optimized on CB1, and the scores were derived based on the dosimetry parameters evaluated on CB2 from Table 5 and scoring functions in Figure 2.

### 3.4 Comparisons of clinical and DT-based treatment plans using daily CBCT images

Following the optimal of plans for Fx1 and Fx2, we assessed the overall treatment performance, as depicted in Figure 1 (Element 11). Table 6 provides a summary of dosimetry comparisons between clinical and DT-based plans for the complete two-fraction prostate SBRT treatment. Notably, for Patient 4, the DT plan enhances CTV V100 coverage by 0.54% and decreases body D0.03cc (hot spot), bladder neck V100 by 1.46%, and 4.08% compared to the clinical plan. For Patient 6, the DT-based plan decreases bladder neck V100 by 42.86% and bladder V14.6Gy by 8.85 cc, compared to the clinical plan. Table 6 also indicates that the scores of the cumulative dose are higher for DT-based treatment regimen. Figures 7(a)-(b) illustrate dosimetry comparisons between the two plans, revealing dose inhomogeneity in the CTV with hot and cold spots for Patient 3 in the clinical plan. Furthermore, Figure 7(c) demonstrates a significant reduction in doses to the bladder neck with the DT-based plan. Similarly, Figures 8(a)-(b) display a cold spot in the CTV and a high-dose region spilling into the bladder in the clinical plan for Patient 7, resulting in higher DVH values in Figure 8(c), whereas the DT-based plan achieves lower doses to the rectum and bladder neck. Lastly, Figures 9(a)-(b) show greater anterior coverage in the clinical plan, leading to high-dose coverage for the bladder neck, consistent with DVH analyses in Figure 9(c) for Patient 8.

**Table 6.** Dose volume endpoint comparisons of CTV and OAR for the ten prostate SBRT patients, obtained by the clinical plans (clinic) and the optimal plans by the proposed digital twin (DT) framework from all fractions. Dose statistics were derived from plan evaluation using CB1 for Fx1 and CB2 for Fx2.

| | | CTV | | Bladder neck | | Body | Bladder | | Rectum | | | Score |
|---|---|---|---|---|---|---|---|---|---|---|---|---|
| | | V100 (%) | D98 (%) | V100 (%) | Dmean (Gy) | D0.03cc (%) | V20.8Gy (cc) | V14.6Gy (cc) | V20.8Gy (cc) | V17.6Gy (cc) | V13Gy (cc) | |
| P01 | Clinic | 97.46 | 99.58 | 10.73 | 18.46 | 105.03 | 4.91 | 11.11 | 0.83 | 1.95 | 4.81 | 165.37 |
| | DT | 97.60 | 99.25 | 7.29 | 17.36 | 104.33 | 3.28 | 8.79 | 0.79 | 1.67 | 4.19 | **167.64** |
| P02 | Clinic | 98.25 | 100.22 | 25.27 | 22.15 | 106.63 | 4.68 | 11.43 | 1.50 | 2.80 | 5.42 | 154.99 |
| | DT | 97.76 | 99.61 | 5.48 | 20.63 | 105.69 | 3.31 | 9.44 | 1.53 | 2.76 | 5.33 | **161.53** |
| P03 | Clinic | 98.59 | 100.42 | 21.17 | 22.47 | 107.76 | 8.33 | 17.29 | 0.70 | 2.16 | 5.20 | 153.08 |
| | DT | 97.70 | 99.68 | 3.89 | 20.38 | 106.03 | 5.63 | 13.81 | 0.64 | 1.88 | 4.58 | **163.30** |
| P04 | Clinic | 99.35 | 100.22 | 33.36 | 16.08 | 107.65 | 13.23 | 22.48 | 0.25 | 1.23 | 4.11 | 152.22 |
| | DT | 99.89 | 100.54 | 29.28 | 15.72 | 106.19 | 14.38 | 24.48 | 0.01 | 0.34 | 2.90 | **157.60** |
| P05 | Clinic | 99.32 | 100.68 | 18.94 | 18.70 | 105.55 | 4.85 | 10.99 | 1.41 | 2.65 | 5.32 | 163.96 |
| | DT | 98.92 | 100.85 | 2.25 | 17.68 | 104.31 | 4.33 | 10.12 | 0.25 | 1.24 | 3.58 | **172.59** |
| P06 | Clinic | 99.92 | 101.48 | 54.76 | 24.30 | 105.67 | 13.89 | 25.79 | 0.35 | 1.36 | 4.06 | 147.38 |
| | DT | 98.59 | 100.50 | 11.90 | 21.53 | 105.77 | 7.47 | 16.94 | 0.17 | 1.00 | 2.94 | **162.91** |
| P07 | Clinic | 99.21 | 100.27 | 31.51 | 18.39 | 105.33 | 17.89 | 29.30 | 0.37 | 1.97 | 6.27 | 151.08 |
| | DT | 99.69 | 101.32 | 9.32 | 16.77 | 104.00 | 12.12 | 22.90 | 0.25 | 1.06 | 4.17 | **165.56** |
| P08 | Clinic | 98.88 | 101.14 | 43.22 | 25.18 | 106.01 | 9.48 | 19.15 | 2.26 | 3.72 | 6.58 | 143.89 |
| | DT | 99.07 | 101.44 | 18.17 | 24.38 | 105.28 | 7.97 | 16.30 | 2.10 | 3.45 | 6.53 | **154.77** |
| P09 | Clinic | 99.85 | 100.63 | 48.04 | 23.45 | 103.95 | 25.55 | 42.07 | 0.89 | 2.18 | 3.60 | 137.61 |
| | DT | 99.69 | 101.26 | 43.32 | 23.31 | 103.54 | 23.88 | 40.38 | 1.08 | 1.95 | 3.94 | **139.62** |
| P10 | Clinic | 99.16 | 100.79 | 9.25 | 10.40 | 106.70 | 16.89 | 29.14 | 1.54 | 3.19 | 6.27 | 154.02 |
| | DT | 99.24 | 101.03 | 9.08 | 9.32 | 106.23 | 12.76 | 24.84 | 0.90 | 2.28 | 5.22 | **163.26** |

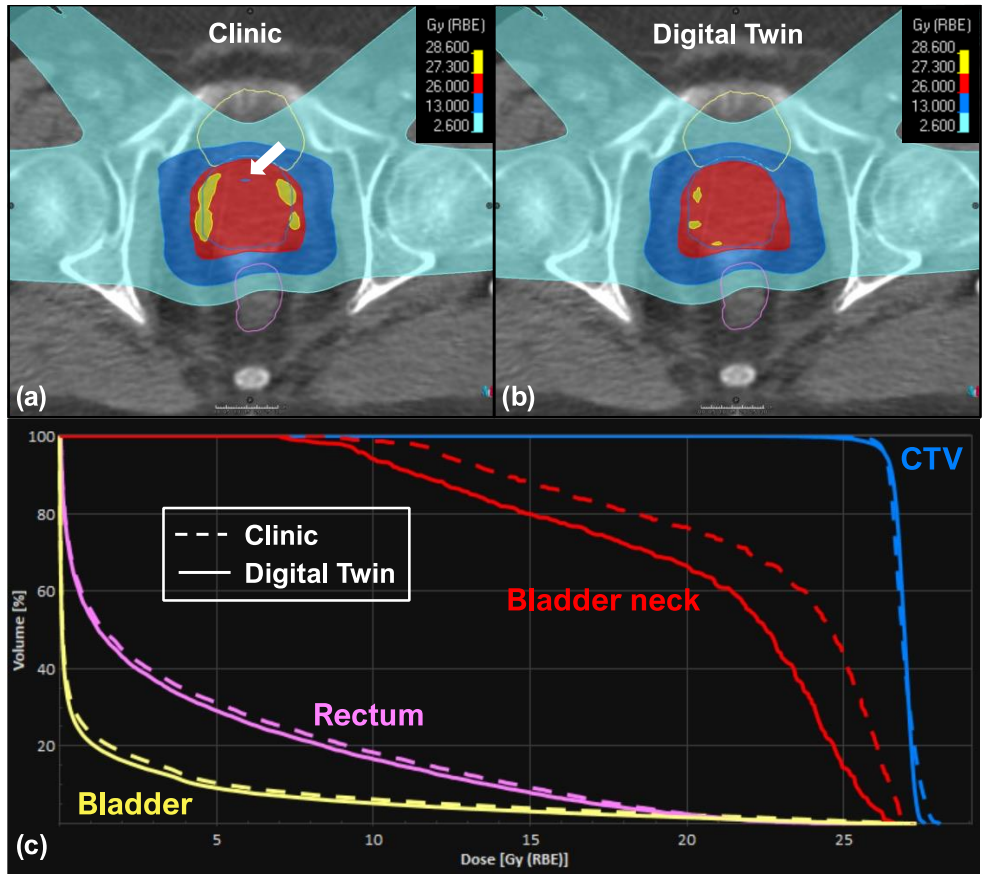

**Figure 7**. Dosimetry comparisons for Patient 3 between the clinical and DT-based plans. The top row displays the dose distributions in color wash for the (a) clinical plan and (b) DT-based plan, overlaid with transversal CT images and the contours of CTV (blue lines), bladder (yellow lines), and rectum (magenta lines). The bottom row (c) displays the dose-volume histogram (DVH) for CTV and OAR structures, where the dashed and solid lines represent the clinical plan and DT-based plan. The white arrow indicates the cold spot location in CTV.

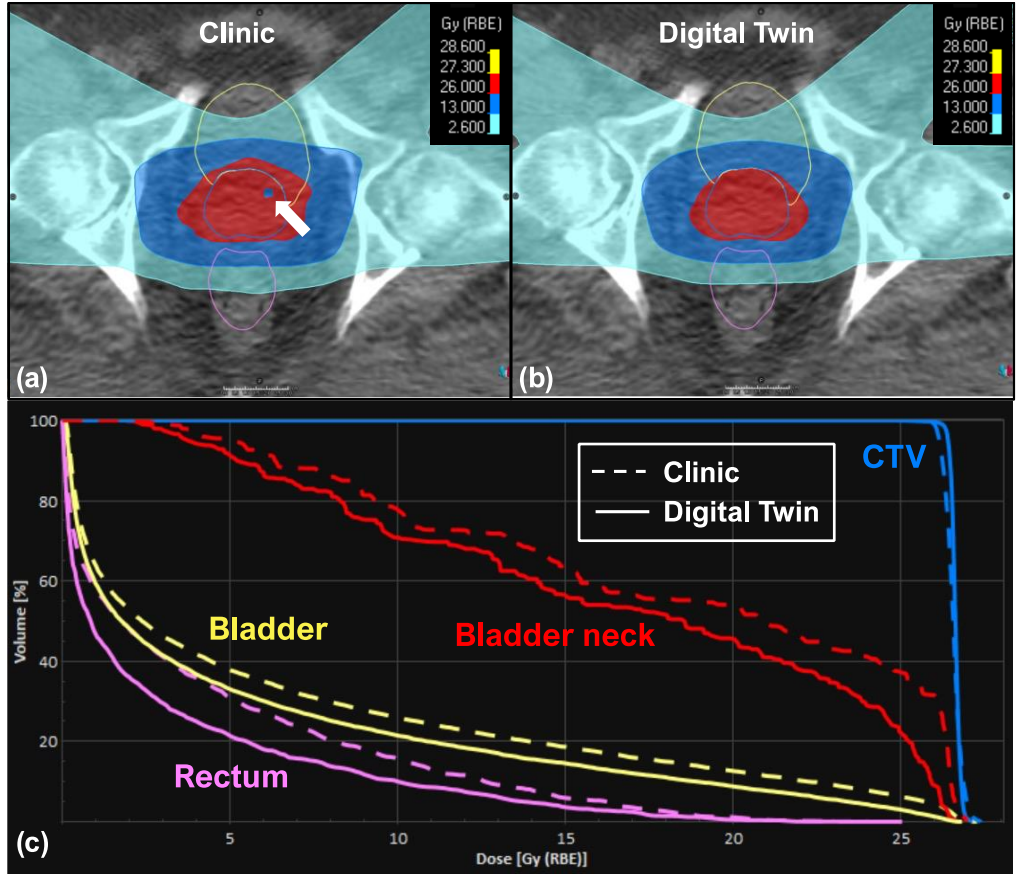

**Figure 8.** Dosimetry comparisons for Patient 7 between the clinical and DT-based plans. The top row displays the dose distributions in color wash for the (a) clinical plan and (b) DT-based plan, overlaid with transversal CT images and the contours of CTV (blue lines), bladder (yellow lines), and rectum (magenta lines). The bottom row (c) displays the dose-volume histogram (DVH) for CTV and OAR structures, where the dashed and solid lines represent the clinical plan and DT-based plan. The white arrow indicates the cold spot location in CTV.

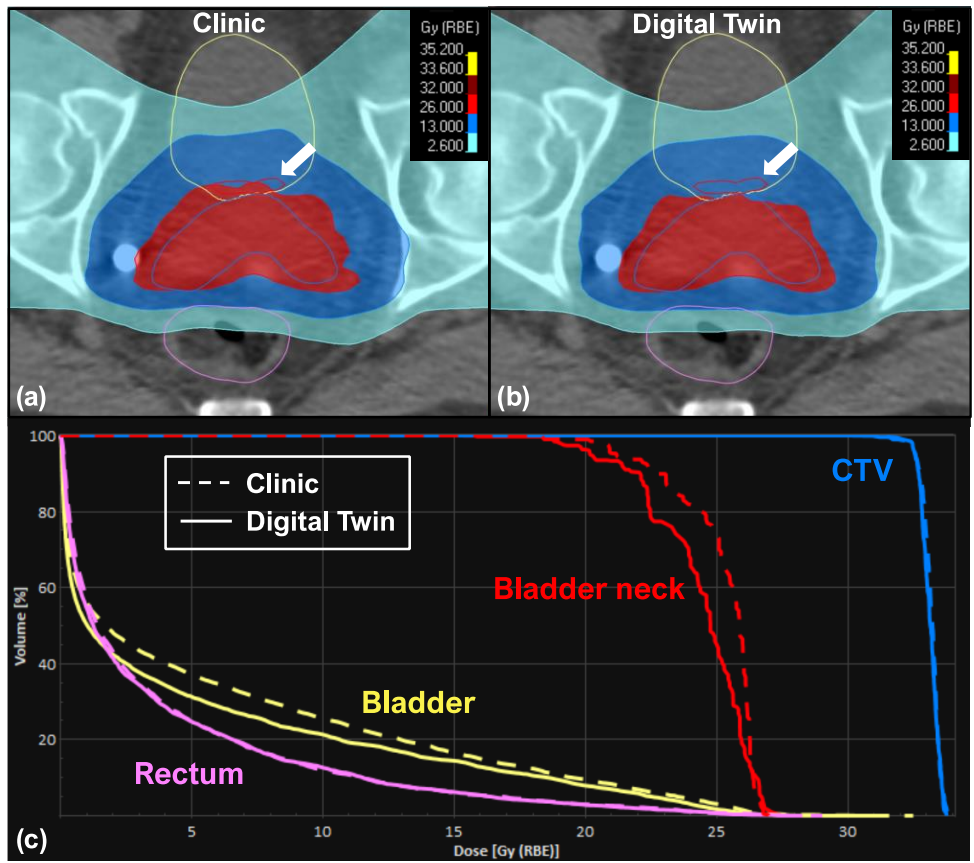

**Figure 9**. Dosimetry comparisons for Patient 8 between the clinical and DT-based plans. The top row displays the dose distributions in color wash for the (a) clinical plan and (b) DT-based plan, overlaid with transversal CT images and the contours of CTV (blue lines), bladder (yellow lines), bladder neck (red lines) and rectum (magenta lines). The bottom row (c) displays the dose-volume histogram (DVH) for CTV and OAR structures, where the dashed and solid lines represent the clinical plan and DT-based plan. The white arrow indicates that the clinical plan results in a significant portion of the bladder neck being covered by the full prescription dose.

## 4 Discussion

The proposed DT framework integrates pre-treatment CT, daily CBCT, patient-specific CTV setup uncertainty, and GPU-based MC TPS to enable the optimal treatment plan selection among the clinic and multiple DT plans. This selection is based on an online plan evaluation using the actual patient anatomy from CBCT at the time of treatment. The patient investigation results demonstrate that the DT plans can improve CTV coverage and significantly reduce bladder neck doses compared to the clinical plan, without the need for re-planning. The current clinical workflow requires re-planning if anatomical changes lead to insufficient CTV coverage or OAR overdoses. The institutional re-planning workflow takes 10 days for a conventional 28-fraction treatment and 5 days for a SBRT treatment. This treatment interruption can potentially impact local control and survival rates for patients (González Ferreira *et al.*, 2015). In contrast, the proposed DT framework pre-generates multiple treatment plans based on the likelihood of potential patient-specific CTV setup uncertainty, supporting on-the-fly treatment decision-making. Compared to current clinical plans, the DT-based plans use smaller patient-specific robust CTV setup uncertainty, resulting in less dose to the surrounding healthy tissue. By addressing the challenges of geometrical uncertainties and optimizing dose conformity, this approach has the potential to improve treatment outcomes, reduce normal tissue toxicity (Prasanna *et al.*, 2021), and ultimately enhance the quality of care for prostate cancer patients undergoing ultra-hypofractionated radiotherapy regimens.

Figures 5-6 illustrate that the majority of DT-based plans achieve higher ProKnow scores compared to clinical plans, suggesting that the current clinical setup uncertainty for CTV might be overly conservative to demonstrate the advantages of proton therapy. Figures 7-9(a)-(b) provide further evidence, showing that the areas covered by prescription doses and the 50% dose color wash are larger for clinical plans due to their wider CTV setup uncertainty compared to DT-based plans. These extended margins of setup uncertainty also pose challenges for the optimizer in finding a global solution during dose optimization. Figure 7(a) reveals dose heterogeneity in the CTV with hot and cold spots. In Figure 8(a), a significant cold spot within the CTV is observed in the clinical plan, potentially increasing the risk of local failures. The DVH comparisons in Figure 8(c) indicate higher doses to the bladder, bladder neck, and rectum compared to the DT-based plan. Figure 9(a) demonstrates that the clinical plan exhibits greater anterior prescription dose coverage, resulting in higher doses to the bladder and bladder neck, as confirmed by the dose color wash and DVH comparisons in Figure 9(c). Table 6 further verifies that DT-based plans yield lower doses to OAR and comparable or higher CTV coverages than clinical plans. Overall, all DT-based plans outperform clinical plans in terms of plan quality according to the ProKnow scoring system.

The proposed DT framework aims at achieving highly precise personalized RT. Two practical inquiries arise to assess the feasibility of this objective: (1) What quantity of candidate treatment plans should be considered? (2) How can the uncertainty in anatomy be effectively measured? Addressing the first query, we advocate for the inclusion of candidate plans from prior treatments in the plan database. Figure 1 illustrates five initial plans, derived from pre-treatment CT (pCT), as potential candidates for treatment administration. The number of candidate plans for Fx2 treatment increases to ten, comprising five plans based on CBCT and five from the initial pCT plans. Figure 6 demonstrates that for Fx2 treatment, 50% of the optimal plans originate from initial pCT plans, while the remaining 50% are based on CBCT plans. Figures 5 and 6(a) display varying optimal plans for patient 9 when assessing the same pCT plans using two distinct CBCT images, CB1 and CB2. For evaluation with CB1, the DT(Fx1) plan emerges as optimal, whereas DT(Fx2)-0.5σ is optimal based on CB2 evaluation. This outcome underscores the importance of incorporating previous candidate treatment plans into the database for future treatment deliberations.

Regarding the issue of anatomy uncertainty, the existing DT framework prioritizes ensuring that the CTV receives prescribed doses while minimizing doses to healthy tissues. To achieve this, GPR serves as a surrogate, consolidating prior knowledge of patient-specific CTV setup uncertainty within potential ranges. As depicted in Table 6, DT-based plans demonstrate the ability to employ smaller CTV setup uncertainty, achieving comparable coverages to clinical plans while delivering reduced doses to OARs and minimizing hot spots. However, the effectiveness of the proposed framework is contingent upon the availability of CBCT images, which represent the patient's actual anatomy on the treatment day, crucial for optimal plan determination. Additionally, the current framework incorporates a minimum 1.5 mm setup uncertainty from the coincidence of radiation and mechanical isocenter centers, this inherent uncertainty that could potentially be reduced with the implementation of more advanced quality assurance techniques for enhanced machine precision. The challenge of proton range uncertainty (Paganetti, 2012), at 3.5%, is another significant consideration stemming from the limitations of CT material characterization methods, potentially compromising the conformity of proton therapy. Future research is likely to explore advanced deep learning-based material density mapping methods (Chang *et al.*, 2022a; Chang *et al.*, 2022b; Chang *et al.*, 2023a) to mitigate this uncertainty and minimize irradiation to normal tissues. A key focus of our future direction is to integrate multi-modality methods including advanced imaging techniques (Hussain *et* *al.*, 2022; Varoquaux and Cheplygina, 2022) and large language models (Bhayana, 2024) into the framework to ensure the level of complexity and data accessibility can enhance the predictive capability of the proposed DT framework.

## 5 Conclusions

We demonstrated a framework harnessing the concept of digital twins to elevate adaptive proton therapy for prostate SBRT. This framework optimized DT-based treatment plans utilizing the most probable patient-specific CTV setup uncertainty, which are smaller than our institutional standard values. Evaluation of plans based on corrected CBCT indicated that DT-based plans achieve superior or comparable CTV coverages while significantly sparing doses to the bladder neck, potentially reducing toxicity risks. By incorporating patient-specific CTV setup uncertainty and integrating prior treatment plans, the proposed DT framework has the potential to assist in treatment decision-making by identifying optimal solutions delivered for radiation oncology patients.

**Acknowledgments**

This research is supported in part by the National Institutes of Health under Award Number R01CA215718, R37CA272755, R01CA272991and R56EB033332.

**Conflict of interest**

The authors have no conflict of interests to disclose.

**Ethical Statement**

Emory IRB review board approval was obtained (IRB #114349), and informed consent was not required for this Health Insurance Portability and Accountability Act (HIPAA) compliant retrospective analysis. The research was conducted in accordance with the principles embodied in the Declaration of Helsinki and in accordance with local statutory requirements.

**Data availability statement**

The data cannot be made publicly available upon publication because they contain sensitive personal information. The data that support the findings of this study are available upon reasonable request from the authors.